\documentclass[10pt,twoside]{hsqcd}
\usepackage{epsf,amsmath}
\usepackage{graphicx}

\begin{document}

\title{Universalities in hadron production and the maximum entropy 
principle}

\author{A.Kropivnitskaya, A.Rostovtsev \\
Institute for Theoretical and Experimental Physics \\
Moscow 117218, Russia\\
E-mail: kropiv@itep.ru, rostov@itep.ru }

\maketitle

\begin{abstract}
A shape of statistical momentum distribution of hadrons produced in high 
energy particle collisions closely resembles one observed for a broad 
variety of phenomena in nature. An attempt was made to understand a 
genesis of this distribution beyond the context of each particular 
phenomenon.
\end{abstract}

A multi-hadron production in high energy hadronic collisions is a 
complicated phenomenon generally modeled by a dynamic system of 
hadronizing quarks and gluons with many degrees of freedom and high level 
of correlations. The properties of produced hadrons at any given 
interaction cannot be predicted. But statistical properties, energy and 
momentum averages, correlation functions, and probability density 
functions show regular behavior. Thus, statistical methods must be applied 
to understand properties of particle production presented by the 
experimentally measured  distributions. 

A statistical model for computing high energy collisions of protons with 
multiple production of particles has been discussed first by E.Fermi in 
1950~\cite{Fermi}. Later on, R.Hagedorn has proposed a statistical 
thermodynamic model to describe the momentum spectra of particles produced 
in $pp-$collisions~\cite{Hagedorn}. This model approximates the 
experimentally measured exponential momentum spectra of hadrons with a 
Boltzmann-like statistical distribution.
With an advance of high energy collision experiments with high statistic 
accumulated the measured momentum spectra are found to deviate from the 
exponential form. Namely, at high values of  particle's transverse 
momentum~$(P_T)$ the spectrum shows a power-law behaviour. This 
observation is interpreted as a proof of the underlying QCD dynamics of 
hadronizing partonic system produced in the particle collisions. Though 
using QCD one calculates elementary interactions of quarks and gluons at 
microscopic level, a complexity of the system generally doesn't allow an 
extension of these calculations to predict the observed particle spectra. 
In addition, it is observed that the relative rates of different particle 
production and parameters of the momentum spectra depend on global 
particle's variables like the mass and spin~\cite{LEP, HERA}. Therefore, 
it is tempting to reconsider a statistical approach to understand the 
particle's transverse momentum spectra in the whole range. 

Firstly, we note that the invariant particle production cross-section is 
approximated by a damped power-law function of a particle's transverse 
momentum:
\begin{equation}
\frac{d\sigma}{dydP_T^2} \sim \frac{1}{(a_0 + P_T)^n}\,,
\label{pdf}
\end{equation}
where $a_0$ and $n$ are parameters of the distribution function. For 
simplicity, we consider only particles produced at central rapidity 
plateau. These particles do not belong to the fragmentation regions of the 
colliding beams.

It is interesting to note, the same form of the damped power-law 
distribution describes a variety of phenomena like a velocity spectrum in 
turbulent liquid, ion energy distribution in geomagnetic plasma,  a 
distribution of the earthquakes as function of it's magnitude (empirical 
Gutenberg-Richter law), distributions of the avalanches and landslides, 
forest fires and solar flares, rains and winds, distributions related to 
the human activities like earnings and settlements sizes, distribution of 
inter-trade intervals observed on stock markets, statistical distribution 
of a number of sexual partners, etc. 
These distributions arise because the same stochastic process is at work, 
and this process can be understood beyond the context of each example. 
Moreover, one can see that for small values of the distributed variable 
the damped power-law expression is reduced to the exponential statistical 
distribution. What is a genesis of the power-law and exponential 
statistical distributions?

    The least biased method to obtain statistical distributions, which are 
realized in the nature was promoted by E.T.Jaynes as Maximum Entropy 
Principle~\cite{MEP}. This Principle states that the physical 
observable has a distribution, consistent with given constraints which 
maximizes the entropy. As an example, the Boltzmann exponential 
distribution arises naturally from a maximization of Gibbs-Shannon entropy 
under a constraint on an average value of the distributed variable. The 
textbook examples of the exponential distributions are the energy of the 
molecules of an ideal gas within an isolated volume and a distance between 
two neighboring points with $N-$points randomly spread over a limited 
interval with length~$L$. Indeed, in both examples the average values of 
the distributed variable are well defined. Namely, due to a conservation 
of the number of molecules and the total energy in the gas volume their 
ratio gives a predefined value of average kinetic energy per molecule of 
gas. 
Similarly, an average distance between two points is obviously defined by 
a ratio of $L/(N-1)$. However, generally speaking, the constraint on the 
average value in most cases has a little sense. Therefore, one needs 
another suitable form of a constraint which results in the damped 
power-law distribution while maximizing the Gibbs-Shannon 
entropy~\footnote{For simplicity here we stick to the Gibbs-Shannon 
entropy form only. Postulating other forms of entropy, like one used in 
non-extensive thermodynamics~\cite{Tsallis} reveals additional conceptual 
problems}. This new constraint has to correspond to a new conservation 
law. There is no obvious known conservation law which could yield the 
damped power-law distribution. Therefore, we propose here a toy model. In 
this model each measurement of a physical value~$x_i$ of interest 
corresponds to a gain of some bits of information equal to $ln(a_0+x_i)$. 
Assume, that an observer is allowed to gain in a set of measurements in 
average a limited information only. This assumption corresponds to the 
following constraint
\begin{equation}
                                \Sigma P_i ln(a_0 + x_i) = const\,,
\label{constraint}
\end{equation}
where $P_i$ is a probability to observe a value $x_i$. Maximization of 
entropy under the constraint~(\ref{constraint}) results in the damped 
power-law probability distribution as function of~$x_i$. It is interesting 
to note that the above constraint~(\ref{constraint}) is reduced to a 
geometric mean value of terms~$(a_0+x_i)$. This makes different events 
correlated with a correlation strength increasing for smaller $a_0$ 
values. A conservation of an average information is lively exemplified in 
a world of fractals. An observer doesn't gain new information about the 
fractal object structure by changing a scale of resolution. If so, a  
meaning of 
the $a_0$ parameter is to define a minimal available information. 
It is interesting to note, the experimental data 
show that for hadron transverse momentum spectra the $a_0$ value is close 
to the mass of produced hadron~\cite{HERA}.  
In addition to the direct measurements of particle spectra a
particle-particle correlation and a broadening of non-Poisson particle 
multiplicity distribution are likely intrinsically related to the 
appearance of power-law tails in particle momentum spectra. Therefore, an 
analysis of their dynamics for varying interaction energies will allow 
further progress. This makes high statistics data accumulated in high 
energy particle interactions to be a unique laboratory to study the whole 
class of the power-law phenomena.

\section*{Acknowledgements} The authors thank the organizing committtee of 
HSQCD'04 
conference for warm hospitality. This work was partially
supported by grants RFBR-03-02-17291 and RFBR-04-02-16445.

\end{document}